\documentclass[paper,11pt]{article}
\usepackage[centertags]{amsmath}
\usepackage{amsfonts} \usepackage{amssymb} \usepackage{amsthm}
\usepackage{jheppub}
\usepackage{tikz}
\usepackage{xparse}
\newcommand{\Z}{{\mathbb Z}}

\newcommand{\one}{{\rm 1\kern -.9mm l}}

\newcommand{\ft}[2]{{\textstyle\frac{#1}{#2}}}

\newcommand{\be}{\begin{equation}}
\newcommand{\eeq}{\end{equation}}
\newcommand{\bea}{\begin{eqnarray}}
\newcommand{\eea}{\end{eqnarray}}
\newcommand{\ba}{\begin{array}}
	\newcommand{\ea}{\end{array}}
\def\nn{\nonumber}

\title{\boldmath The chiral ring of gauge theories in eight dimensions}
\author[a]{F. Fucito}
\author[a]{J.F. Morales,}
\author[b]{and R.Poghossian}
\affiliation[a]{I.N.F.N - sezione di Roma 2,\\and Universit\`a di Roma Tor Vergata, Dipartimento di Fisica\\Via della Ricerca Scientifica, I-00133 Roma, Italy}
\affiliation[b]{Yerevan Physics Institute,\\Alikhanian Br. 2, AM-0036 Yerevan, Armenia}

\vspace{0.35cm}
\emailAdd{fucito@roma2.infn.it}
\emailAdd{morales@roma2.infn.it}
\emailAdd{poghos@yerphi.am}

\abstract{ We study the non-perturbative corrections generated by exotic instantons in U(N) gauge theories in eight and four dimensions. 
	As it was shown previously, the eight-dimensional prepotential can be resummed using a plethystic formula showing only a dependence from the center of mass and from a
	U(1) gauge factor. On the contrary, chiral correlators in eight and four dimensions  display a non-trivial dependence from the full gauge group. Furthermore the resolvent, the generating function for the eight and four dimensional correlators, can be written in a compact form both in the eight and four dimensional cases.
	
}


\begin{document}
	\maketitle
	\flushbottom
\section{Introduction}
\label{secn:intro}

In recent years the study of D{\it p}-D{\it p}$^\prime$ systems has been particularly fruitful. Such systems can form bound states and preserve part of the original supersymmetry: they are BPS states. 
Of particular interest is the D{\it p}-D({\it p}+4) system which accounts for non perturbative effects in four dimensional theories reproducing the ADHM construction for the moduli space of four dimensional gauge instantons \cite{Witten:1995im,Douglas:1996uz,Green:1997tn,Green:1998yf,Green:2000ke,Billo:2002hm}. In the case in which eight supersymmetric charges are conserved, the prepotential of the four-dimensional gauge theory is codified in the free energy of the statistical partition function counting the low energy excitations of open strings connecting the various D-branes. 

The computations of the partition function and of the chiral ring of these theories can be
carried out using localization. 

Gauge theories with gauge groups $U(N)$ are realised in term of oriented open strings while $SO(N)$ and $Sp(N)$ gauge groups arise when the brane system is on top of an orientifold plane O{\it p}$^\prime$.  
These considerations  extend to the case in which the two sets of branes have eight directions with different boundary conditions. 
The spectrum of twisted non oriented open strings in this case exhibits a lack of bosonic moduli  associated to the instanton sizes and gauge orientations. As a consequence these systems exhibit prepotentials behaving as an infinite series of exponentially suppressed corrections in the weak coupling with coefficients given by finite polynomials in the Casimirs formed with the expectation values of the scalar fields in the Coulomb branch. This is on the contrary of what happens in the four dimensional case where, in the weak coupling limit, the coefficients of the prepotential series are given by the inverses of the above mentioned Casimirs.
 These non perturbative effects, triggered by point like branes dubbed as ``exotic instantons", 
are of great phenomenological interest for their ability to generate Majorana mass terms or Yukawa couplings  in four dimensional theories obtained from the eight dimensional theory by dimensional reduction and orbifolding.  The four-dimensional counterpart of these exotic instantons have been extensively studied with the same motivations \cite{Blumenhagen:2006xt,Ibanez:2006da,Argurio:2006ny,Argurio:2007vqa,Bianchi:2007wy,Blumenhagen:2007zk,Ibanez:2007rs,Ibanez:2007tu}.   Applications to holography in D(-1)-D3-D7 systems has been considered in  \cite{Billo:2010mg,Fucito:2011kb}.

Recently the oriented version of the D(-1)-D7 system (or equivalently D0-D8) plus matter has been studied from another perspective \cite{Nekrasov:2017cih,Nekrasov:2018xsb}. In absence of the orientifold plane, this system can be made stable by turning on fluxes on the D7-branes so that the ground state of the NS sector comes down to zero mass. The spectrum of massless open strings now has a bosonic degree of freedom and describes the eight-dimensional analog of the ADHM construction of $U(N)$ instantons \cite{Billo20201}. The system can be further decorated by adding $M$ flavour D7 branes with trivial worldvolume fluxes.  The gauge partition function for $M=N$ has been computed in \cite{Nekrasov:2017cih,Nekrasov:2018xsb} and written in a compact form using a plethystic resummation. Quite surprisingly the final result is only dependent on an overall $U(1) \in U(N)\times U(M)$ subgroup of the original gauge group.  Moreover, the result vanishes in the flat space limit where the $\Omega$ background used by localization is turned off. This means that non perturbative contributions in this theory modify the prepotential only in presence of gravitational fields. 

A complete analysis of this brane system from the point of view of string theory will appear in \cite{Billo20201}.
The aim of this paper is to study the chiral correlators (the chiral ring) of the theory. The results show that unlike the gauge theory prepotential, chiral correlators display a non-trivial dependence on the full gauge group $U(N)$. Again, non-trivial results require the presence of gravitational fields and can be resummed to all instanton orders in closed form.
This is the plan of the paper: in Section 2 we recall some basic elements of our construction. In Section 3 we compute the resolvent in the eight dimensional case and finally in Section 4 we extend our computations to the four dimensional case. In the appendix we collect some results concerning the transformation properties of the eight dimensional resolvent under
the $SL(2,\Z)$ group.   
\section{The eight dimensional theory}

In this section we review the basic elements of our theory and we introduce the character which allows to compute the prepotential and the chiral correlators generated by exotic instantons in the $U(N)$ eight-dimensional gauge theories built out of D7-branes with non-trivial worldvolume fluxes.

\subsection{D(-1)-D7 system on ${\bf R}^8$ }

We consider a system of $k$ D(-1) branes, $N$ D7-branes with a non-trivial worldvolume flux and $M$ flavour D7'-branes (see \cite{Billo20201} for details on the brane construction of the system).
 The instanton moduli arise from the open strings connecting the various branes and with at least one end on the D(-1)-branes.
 The moduli and the effective action of the theory coming from the open strings starting and ending on $k$ D(-1) branes can be obtained by reducing to zero dimension the $U(k)$ gauge theory (SYM) with $N=1$ supersymmetry in ten dimensions. 
The bosonic field content includes four complex scalars $B_\ell$, $\ell=1,2,3,4$ coming from the gauge vector components along the D7, and a complex field $\chi$ from the transverse components.  The 32 components of the spinor in ten dimensions are reduced to 16 because our system is 1/2 BPS. In turn these sixteen components, which from the point of view of group theory are in the representations $8_s+8_c$, will give the supersymmetric partners of the $B_i$ and a fermionic lagrangian multiplier to implement the fermionic constraint on the moduli. Under SU(4) we will get $8_s=4\oplus\bar 4$ and $8_s=6\oplus 1\oplus 1$ so we will describe these degrees of freedom as four complex fermions $M_\ell$, the partners of $B_\ell$, a component $\lambda_-$ paired under supersymmetry with
$\bar\chi$ and seven real components  $\{\lambda_{[IJ]}, \lambda^{[IJ]},\lambda_+ \}$  with $I,J=1,2,3$ and $I<J$
 paired with the seven auxiliary fields
$\{D_{[IJ]}, D^{[IJ]},D_7 \}$ implementing the vacuum conditions.  All D(-1)-D(-1) fields transform in the adjoint of the $U(k)$ symmetry group. 
 On the other hand the D(-1)-D7 and D(-1)-D7' strings provide the moduli $(w,\mu)$ and $(\mu',h)$ in the $U(N)\times U(k)$ and $U(M)\times U(k)$ bifundamental representations respectively.
 
To localize the partition function we have to make the theory equivariant with respect to the Cartan of the group $G=U(k) \otimes U(N)\otimes U(M) \otimes SO(8)$ where $U(k)$, $U(N)$, $U(M)$ and $SO(8)$ are the instanton, gauge, flavour and eight-dimensional Lorentz groups respectively. We parametrize the Cartan subgroups by $\chi={\rm diag} \{\chi_i \}$, $a={\rm diag} \{a_u \}$, $m={\rm diag} \{m_s \}$ and $\epsilon_\ell$ respectively with
\begin{equation}
\epsilon_1+\epsilon_2+\epsilon_3+\epsilon_4=0 \label{ezero}
\end{equation}
given that the $SO(8)$ is broken to $SO(7)$ since there are seven ADHM constraints for our system.
The diagonal matrices $\chi$, $a$ and $m$ parametrize the positions of the D(-1), D7 and D7' branes respectively along the direction  transverse to the D7-branes. The $a_u$'s describe the vev's of the scalar fields and the $\epsilon_\ell$'s the eight-dimensional $\Omega$-background.

\begin{table}[t]
\begin{center}
\begin{tabular}{|c|c|c|c|}
\hline
$(\Phi,{\cal Q} \Phi )$ & $(-)^F$ &$U(k)\times U(N)\times U(M)$ & ${\cal Q}^2$ \\
\hline
$(B_\ell,M_\ell )$ & $ + $ &  $  ({\bf k} {\bf \bar k} , {\bf \one}, {\bf \one} )$ & $T_\ell \, {x_i\over x_j} $ \\
$(\bar B_\ell,M^\ell )$ & $ + $ &  $  ({\bf k} {\bf \bar k} , {\bf \one}, {\bf \one} )$ & $T_\ell^{-1} \, {x_i\over x_j} $ \\
$(\lambda^{IJ},D^{IJ} )$ & $-$ & $  ({\bf k} {\bf \bar k} , {\bf \one}, {\bf \one} )$ & $T_I T_J \, {x_i\over x_j} $ \\
$(\lambda_{IJ},D_{IJ} )$ & $-$ & $  ({\bf k} {\bf \bar k} , {\bf \one}, {\bf \one} )$ & $T_I^{-1} T_J^{-1} \, {x_i\over x_j} $ \\
$(\lambda_+,D_7 )$ & $-$ & $  ({\bf k} {\bf \bar k} , {\bf \one}, {\bf \one} )$ & $  {x_i\over x_j} $ \\
$(\bar\chi,\lambda_- )$ & $+$ & $  ({\bf k} {\bf \bar k} , {\bf \one}, {\bf \one} )$ & $ {x_i\over x_j} $ \\
$(w,\mu)$ & $+$ & $  ({\bf \bar k}  , {\bf  N}, {\bf \one} )$ & ${T_{a_u}\over x_i} $ \\
$(\mu', h')$ & $-$ & $  ({\bf \bar k} , {\bf \one} , {\bf M} )$ & ${T_{m_f}\over x_i} $ \\
\hline
\end{tabular}
\end{center}
\caption{Instanton moduli space. The domain of the indices for $\mathbb{R}^8$ are: $\ell=1,\ldots 4$, $I,J=1,\ldots 3$ with $I<J$, $i, j=1,\ldots,k$, $u=1,\ldots,N$ and $f=1,\ldots,M$.   }
\label{tablemoduli8}
\end{table}%
   
  The equivariantly deformed supersymmetric variation ${\cal Q}$ reads
\bea
{\cal Q} \, B_\ell &=& M_\ell \qquad , \qquad {\cal Q}\, M_\ell=[\chi ,B_\ell]  +\epsilon_\ell\, B_\ell\nn\\
{\cal Q} \, \lambda^{IJ} &=& D^{IJ} \qquad , \qquad {\cal Q}\, D^{IJ}=[\chi ,\lambda^{IJ}]  +(\epsilon_I+\epsilon_J) \lambda^{IJ}\nn\\
{\cal Q} \, \lambda_{+} &=& D_7 \qquad , \qquad {\cal Q}\, D_7=[\chi ,\lambda_{+}]  \nn\\
{\cal Q} \, \bar \chi  &=& \lambda_- \qquad , \qquad {\cal Q}\, \lambda_-=[\chi ,\bar \chi]  \nn\\
{\cal Q} \, w &=& \mu \qquad , \qquad {\cal Q} \mu=\chi w +w\, a\nn\\
{\cal Q} \, \mu' &=& h \qquad , \qquad {\cal Q} h=\chi \mu' + \mu' \, M\nn\\
{\cal Q} \,  \chi  &=& 0
\eea
 The field content and the transformation properties with respect to  $U(k)\times U(N)\times U(M)\times SO(8)$  for the eight dimensional case are summarized 
 in Table \ref{tablemoduli8}. We find it convenient to compute the ${\cal Q}^2$ eigenvalues of the various fields by introducing the following character
\bea 
\label{char} 
\chi ={\rm tr}_{\cal H}  e^{-\beta {\cal Q}^2 } ={-}\sum_{i<j=1}^k\frac{x_i}{x_j}\prod_{\ell=1}^4(1{-}T_\ell) {-}  \sum_{i=1}^k \prod_{I=1}^3 (1{-}T_I)  
+\sum_{i=1}^k \left( \sum_{u=1}^N { T_{a_u}  \over x_i } {-}   \sum_{f=1}^M   { T_{m_f}  \over x_i}  \right) 
\eea
where
\be
T_\ell =e^{-\beta \epsilon_\ell} \quad, \quad   
x_i =e^{-\beta \chi_i} \quad, \quad   T_{a_u} =e^{-\beta a_u} \quad, \quad   
 T_{m_f} =e^{-\beta m_f} 
\eeq
According to (\ref{ezero}) one finds that the $T_\ell$'s satisfy the relation
\be
T_1T_2T_3T_4=1
\label{epsilonconstraint9d}
\eeq
The character (\ref{char}) provides the natural generalization of the four dimensional character which, as detailed in \cite{Fucito:2004gi}, keeps into account the contributions of the various degrees of freedom.
 Each monomial in (\ref{char}) represents a supermultiplet in Table \ref{tablemoduli8} contributing with plus or minus depending on whether the multiplet involve a physical or an auxiliary field respectively. In particular the linear and cubic terms in $T_\ell$ in the expression 
 $ \prod_{\ell=1}^4(1{-}T_\ell) $ represent the contributions of $B_\ell$, $\bar B^\ell$, while the quadratic ones subtract the 
 degrees of freedom associated to the 
 ADHM constraints ( $D_{IJ}$ and $\bar{D}^{IJ}$).  
  The novelty with respect to the four dimensional case is that, due to the constraint (\ref{epsilonconstraint9d}),  the term $\prod_{\ell=1}^4(1{-}T_\ell)$  in the eight dimensional character contains not only the degrees of freedom of the fields but also the contribution of their complex conjugates thus doubling the degrees of freedom. To restore the right counting of degrees of freedom we restrict the sum over $i,j$ to $i<j$. The second term in (\ref{char}) subtracts half the degrees of freedom of the diagonal $i=j$ terms. Finally the last two terms in (\ref{char}) account for the open strings ending on D7 and D7' branes.

After localizing and computing  the determinant ${\rm det} {\cal Q}^2$, the partition function is given by  $Z_{9d}=\sum_k Z_{k} \, q^k$
with
 \footnote{The terms appearing in the integrand of (\ref{localization}) come straighforwardly from the character (\ref{char}). Each term $\bullet$ in the character contributes a (1-$\bullet)^{\pm}$  with $\pm$ depending on the sign of the monomial in the character. } 
\bea
Z_k &=&    {{\cal E}^k\over k!} \int  \prod_{i=1}^k {dx_i\over x_i}  \prod_{i < j}^k\frac{ \left(1-{x_i \over x_j} \right)  \left( 1-{x_i \over x_j} T_1 T_2 \right) \left( 1-{x_i \over x_j} T_1 T_3 \right)
 \left( 1-{x_i \over x_j} T_2 T_3 \right) }{  \left( 1-{x_i \over x_j} T_1\right)\left( 1-{x_i \over x_j} T_2\right)\left( 1-{x_i \over x_j} T_3\right)\left( 1-{x_i \over x_j} T_4\right)} \nn\\
 && \times  \prod_{i=1}^k \frac{\prod_{f=1}^M \left(1-{T_{M_f} \over x_i} \right)} {\prod_{u=1}^N  \left(1-{T_{a_u} \over x_i} \right)  }
\label{localization}
\eea
with
\be
{\cal E}= \frac{ \left( 1-  T_1 T_2 \right) \left( 1- T_1 T_3 \right) \left( 1-  T_2 T_3 \right) }{  \left( 1-  T_1\right)\left( 1-  T_2\right)\left( 1- T_3\right) \left( 1- T_4\right)}
\eeq
We refer to $Z_{\rm 9d}$ as the nine dimensional partition function, resulting from the lift of the eight dimensional theory to $\mathbb{R}^8\times S^1$ along a circle of radius $\beta$. 
The case $N=M$ was the one considered in \cite{Nekrasov:2018xsb}.
Formula (\ref{localization})  has some sign differences from the ones appearing in \cite{Nekrasov:2018xsb} where these signs have been fixed 
order by order to produce the plethystic result displayed below. We have checked that the partition function computed  using (\ref{localization}) reproduces the plethystic exponential up to $k=15$ without any sign addition.

The poles of the integrand in (\ref{localization}) can be put in correspondence with sets of $N$ solid partitions (four dimensional Young diagrams) with $k$ total number of boxes (instanton number). Imposing a lexicographic order we numerate the boxes of the set 
$ \{ Y_u \}$ by a single integer label $i$, each associated to a box with coordinates $(m_1,m_2,m_3,m_4)\in Y_u$ and
coordinate $x_i^Y$ given by  
\be
\sum_{i=1}^k x^Y_i =\sum_{u=1}^N\sum_{\{ m_\ell \} \in Y_u}^kT_{a_u} T_1^{m_1-1}T_2^{m_2-1}T_3^{m_3-1}T_4^{m_4-1}
\label{Vspace}
\eeq
 The partition function is given by the sum $Z_{9d}=\sum_Y Z_{9Y} \, q^{|Y|}$ over the $N$ solid partitions $\{ Y_u \}$ of the residues of the integrand in (\ref{localization}) 
 at $x_i=x_i^Y$. 
 For instance for $k=1$ one finds, N diagrams contributing, each with a single box centered at $a_u$. 
 If $N=M$ after summing over the $N$ contributions one finds
 \be
 Z_1=   \frac{ \left( 1-  T_1 T_2 \right) \left( 1- T_1 T_3 \right) \left( 1-  T_2 T_3 \right) (1-T_\mu)}{  \left( 1-  T_1\right)\left( 1-  T_2\right)\left( 1- T_3\right) \left( 1- T_4\right)}  
 \eeq
with $T_\mu=e^{-\beta \mu }$ and
\be
 \mu=\sum_{u=1}^N \left(m_u-a_u \right) \label{mu}
\eeq
%In the case $M=N-1$ 
%\be
%Z_1=   \frac{ \left( 1-  T_1 T_2 \right) \left( 1- T_1 T_3 \right) \left( 1-  T_2 T_3 \right)}{  \left( 1-  T_1\right)\left( 1-  T_2\right)\left( 1- T_3\right) \left( 1- T_4\right)}  
%\eeq
%and if the number of matter multiplets is even less i.e. $M\le N-2$, one simply gets $Z_1=0$.
  Similarly for higher $k$, the partition function can be computed. Remarkably, the all instanton result can be written in the simple plethystic form  \cite{Nekrasov:2018xsb}
  \be
  Z_{\rm 9d}={\rm exp} \left[  \sum_{n=1}^\infty { f(T^n_\ell, T^n_\mu,q^n) \over n } \right] \label{plethystic} 
  \eeq 
where, as usual, $q=e^{2\pi i\tau}$. $\tau$ is the complex coupling of our gauge theory and
\be
f(T_\ell, T_\mu,q) = \frac{\left( 1-  T_1 T_2 \right) \left( 1- T_1 T_3 \right) \left( 1-  T_2 T_3 \right) q}{  \left( 1-  T_1\right)\left( 1-  T_2\right)\left( 1- T_3\right) \left( 1- T_1T_2T_3\right)} {(1-T_\mu)\over (1-q T_\mu)(1-q) } \label{fple}
\eeq
%if $M=N$ and
%\be
%f(T_\ell, T_\mu,q) = \frac{\left( 1-  T_1 T_2 \right) \left( 1- T_1 T_3 \right) \left( 1-  T_2 T_3 \right)}{  \left( 1-  T_1\right)\left( 1-  T_2\right)\left( 1- T_3\right) \left( 1- T_1T_2T_3\right)} {q\over (1-q) }
%\eeq
Remarkably, the results either depend only on the specific combination $\mu$ involving the total mass and the overall center of mass $U(1)$ .  The results for $M<N$ can be obtained by sending  $\mu \to  \infty$, leading again to the plethystic formulas 
(\ref{plethystic}-\ref{fple})  with $T_\mu= 0$.  Remarkably all dependence on the remaining masses and expectation values cancel in this limit. 
  
%if $M=N-1$. In all remaining cases $M\le N-2$ we get $f(T_\ell, T_\mu,q) =0$ i.e. $Z_{\rm 9d}=1$.  
%when $M=N$ or,  in remaining cases, do not depend on these parameters whatsoever.

 The eight dimensional  partition function is obtained from $Z_{9d}$ by sending $\beta$ to zero
 \be
  Z_{\rm 8d} (a,m,\epsilon_\ell,q)={\rm lim}_{\beta\to 0}  Z_{\rm 9d}(T_a,T_m,q)
  \eeq 
Again the partition function can be written as a sum over 4d solid partitions $\{ Y_u \}$  
 \be
  Z_{\rm 8d} (a,m,\epsilon_\ell,q)=\sum_k q^k \, Z_{8k}= \sum_{Y} q^{|Y|} \, Z_{8Y}
  \eeq

\bea
Z_{8k} &=&    {{\cal V}^k\over k!} \int  \prod_{i=1}^k {d\chi_i}  \prod_{i < j}^k\frac{ \chi_{ji}   \left( \chi_{ji}+\epsilon_1+\epsilon_2 \right)  \left( \chi_{ji}+\epsilon_1+\epsilon_3 \right)  \left( \chi_{ji}+\epsilon_2+\epsilon_3 \right) 
  }{    \left( \chi_{ji}+\epsilon_1 \right)   \left( \chi_{ji}+\epsilon_2 \right)    \left( \chi_{ji}+\epsilon_3 \right)    \left( \chi_{ji}+\epsilon_4 \right)  }
   \nn\\
 && \times 
 \prod_{i=1}^k \frac{\prod_{f=1}^M \left( \chi_i - m_f \right)} {\prod_{u=1}^N  \left(\chi_i-a_u\right)  }
\label{localization2}
\eea
with $\chi_{ji}=\chi_j-\chi_i$, and
\be
{\cal V}={\left( \epsilon_1+\epsilon_2 \right)  \left(  \epsilon_1+\epsilon_3 \right)  \left( \epsilon_2+\epsilon_3 \right) \over
\epsilon_1\epsilon_2\epsilon_3\epsilon_4}
\eeq
The eight-dimensional prepotential is defined as
 \bea
\label{logZ8d}
{\cal F}_8 (a,\epsilon_\ell,q)=\epsilon_1\epsilon_2\epsilon_3\epsilon_4  \log Z_{8d}
\eea

In the case $M=N$ we  get
\bea
\label{logZ8d1}
{\cal F}_8 (a,\epsilon_\ell,q)=(\epsilon_1+\epsilon_2)(\epsilon_1+\epsilon_3)
(\epsilon_2+\epsilon_3) \mu \sum_{n=1}^\infty  \frac{q^n}
{n (1-q^n)^2}\nn\\
=(\epsilon_1+\epsilon_2)(\epsilon_1+\epsilon_3)
(\epsilon_2+\epsilon_3)  \,\mu\,\log \mathcal{M}_3(q)
\eea
where
\be 
\mathcal{M}_3(q)=\prod_{n=1}^\infty \frac{1}
{(1-q^n)^n}
\eeq 
is the Mac Mahon function counting the three dimensional partitions of the integers. The factor $\epsilon_1\epsilon_2\epsilon_3\epsilon_4$ 
gets rid of the superspace volume  in eight dimensions $\int d^8 x d^8 \theta=1/(\epsilon_1\epsilon_2\epsilon_3\epsilon_4)$.
We notice that formula (\ref{localization2}) for generic $M$ can be obtained from that for $M=N$ by sending some masses to infinity
keeping constant $\hat{q}=q \prod_{f>M}^N (-m_f)$.  
The results for $M<N$ can then be obtained from (\ref{logZ8d1}) by sending $\mu\to \infty$ and $q\to 0$ with $\mu^{N-M} q$ kept fixed. 
One finds
\bea
\label{logZ8d2}
{\cal F}_8 (a,\epsilon_\ell,q)=
\left\{
\begin{array}{cc}
 (\epsilon_1+\epsilon_2)(\epsilon_1+\epsilon_3)
(\epsilon_2+\epsilon_3) \hat{q}  &  \qquad     M=N-1     \\
   0 & \qquad M<N-1
\end{array}
\right.
\eea 
 The prepotential ${\cal F}_8$ defines the non-perturbative effective action of the eight-dimensional theory
 \be
S_{\rm eff}=\int d^8xd^8\theta\, {\cal F}_8(\Phi,W,q) %=\tau(a,\epsilon)\int \,d^4x \,tr F^2+\ldots
\label{effaction}\eeq
after the parameters $a$ and $\epsilon_\ell$ are promoted to superfields $\Phi, W$.   We observe that  
only the $U(1)$ part of the prepotential is affected by instanton corrections. We will show below that this is not the case for the correlators of the gauge theory that exhibit a full dependence from the $SU(N)$ gauge degrees of freedom.

\subsection{Chiral correlators}

In this section we will focus on the chiral correlators $\langle {\rm tr}\phi^J \rangle$ of the gauge theory. 
According to localization \cite{Fucito:2009rs}, the generating function of these correlators is given by  
\be
\label{trfij}
\langle {\rm tr}\, e^{ z\, \phi} \, \rangle = {\rm tr}\, e^{ z\, a}- {1\over Z_{8d} }  \sum_{Y} q^{|Y|}\, Z_{8Y}    \sum_{i=1}^k \left(x_{i}^Y \right)^z   \prod_{\ell=1}^4 (1-T_\ell^z) 
\eeq
  $x_{i}^Y$ denotes the location of the corresponding pole given by (\ref{Vspace}).  We notice that in the eight-dimensional limit $\beta \to 0$, the right hand side of (\ref{trfij}) starts as $z^4$, so all correlators with $J<4$ receive no instanton corrections. The first non-trivial correlator is then given by the coefficient of the $z^4$
term in (\ref{trfij})
\be
\label{trfij4}
\langle {\rm tr}\,   \phi^4 \, \rangle = {\rm tr}\, a^4- {24\epsilon_1\epsilon_2\epsilon_3\epsilon_4 \over Z_{8d}}  \sum_{Y} k\,q^k Z_{8k}   
=   {\rm tr}\, a^4 -24q{d\over dq} {\cal F}_8(q)
\eeq
This provides the eight-dimensional analog of the relation found in \cite{Fucito:2009rs}.
  Similar higher correlators can be evaluated order by order in $q$. 
We find it convenient to pack the correlators $\langle {\rm tr}\phi^J \rangle$ into a different generating function, the so called resolvent, defined by
\be
\left\langle tr\,\frac{1}{x-\phi} \right\rangle=\sum_{J=0}^{\infty}
\frac{\langle tr\, \phi^J \rangle}{x^{J+1}}
\eeq   
This function is closely related to the Seiberg-Witten curve \cite{Nekrasov:2003rj} and will be written below in a closed form.

Let us consider first the $U(1)$ case without mass multiplets, i.e. $N=1$, $M=0$. Up to order $q^{16}$ one finds
\bea
&&\left\langle tr\,\frac{1}{x-\phi} \right\rangle =\frac{1}{x-a} +12\epsilon^{(3)}\left( 
\frac{2 q}{(x-a)^5}-\frac{25 q^2}{(x-a)^6}+\frac{100 q^3}{(x-a)^7}
-\frac{735 q^4}{2 (x-a)^8}\right.
\nonumber\\
&& \left. +\frac{728 q^5}{(x-a)^9}
-\frac{2100 q^6}{(x-a)^{10}}  +\frac{3000 q^7}{(x-a)^{11}}
-\frac{14025 q^8}{2 (x-a)^{12}}+\frac{10010 q^9}{(x-a)^{13}}
-\frac{18590 q^{10}}{(x-a)^{14}} \right. \\
&&
\left. +\frac{22204 q^{11}}{(x-a)^{15}}
-\frac{47775 q^{12}}{(x-a)^{16}}+\frac{47600 q^{13}}{(x-a)^{17}}
-\frac{85000 q^{14}}{(x-a)^{18}}
+\frac{106080 q^{15}}{(x-a)^{19}}
-\frac{330429 q^{16}}{2(x-a)^{20}}+\ldots
\right)+O\left(\epsilon^4\right) \nn
\eea
 with
\be
\epsilon^{(3)}\equiv (\epsilon_1+\epsilon_2)(\epsilon_1+\epsilon_3)(\epsilon_2+\epsilon_3).
\label{epsilon3}\eeq
After some educated guessing the full answer can be written in closed form as
\be
\label{resolv_exact} 
\left\langle tr\,\frac{1}{x-\phi} \right\rangle =\frac{1}{x-a}-
\epsilon^{(3)}
\frac{\partial ^3}{\partial x^3}\left[ \frac{1}{a-x} D_3 \left(\frac{q}{a-x}\right)\right]+O\left(\epsilon^4\right)
\eeq 
where we introduce the function  
\be
D_3(q)=\sum_{n=1}^{\infty}\sigma_2(n)q^n=\sum_{k=1}^{\infty}\frac{k^2 q^k}{1-q^k} =q{d\over dq} \log{\cal M}_3(q) 
\label{divisors2}\eeq 
with
\be
\sigma_2(n)=\sum_{d|n}d^2
\eeq 
the sum of all squared divisors $d^2$ of $n$. The generalization to $U(N)$ with flavour group $U(M)$ is straightforward. 
 One finds
\be
\label{resolv_exact_mass} 
\left\langle tr\,\frac{1}{x-\phi} \right\rangle =\frac{P'(x)}{P(x)}-
\epsilon^{(3)}
\frac{\partial ^3}{\partial x^3}
\left[ \frac{R'(x)}{R(x)}
D_3\left(qR(x)\right) +O(\epsilon)\right]
\eeq 
with 
\be
\label{P_Q_def}
P(x)=\prod_{u=1}^N(x-a_u)\,    \qquad ,\qquad  R(x)=\frac{\prod_{f=1}^M(m_u-x)}{\prod_{u=1}^N(a_u-x)}\,
\eeq  
Once again we notice that, as in the case of the prepotential, these correlators get corrected only in the presence of gravitational fields given the presence of the factor (\ref{epsilon3}), but now non-trivial instanton corrections modify also the $SU(N)$ part of the theory. 
For instance, specifying to $SU(2)$ theory with two flavours, and taking $a_1=-a_2=a$, $m_1=-m_2=m$, one finds for the first non-trivial 
correlators
\bea
\langle {\rm tr}\,   \phi^5 \, \rangle &=& 120 \epsilon^{(3)} (m^2-a^2) ( q+ 5 q^2+10 q^2+\ldots )\nn\\
 \langle {\rm tr}\,   \phi^7 \, \rangle &=& 420 \epsilon^{(3)} (m^2-a^2) ( 2q+ 5 (3a^2-m^2) q^2+ 20 (2a^2-m^2) q^3+\ldots ) 
\eea
 and so on.
We conclude this section observing that the infinite series $D_3(\tau)$ appearing in the correlators differs from an Eisenstein series because the power of $k$ in (\ref{divisors2}) is even rather than odd. This difference modifies drastically the behaviour of 
the function under $SL(2,\Z)$ transformations of its gauge coupling $\tau$. We find (see Appendix for details)
\bea\label{d3transf}
&&D_{3}\left(\frac{a\tau+b}
{c\tau+d}\right)=\left(c\tau+d\right)^{3}D_{3}\left(\tau\right)\nonumber\\
&&+\frac{i}
{4\pi^3}\left(\left(1+(c\tau+d)^{3}\right)
\zeta(3)+2\sum_{m=1}^\infty\sum_{n=1}^\infty 
\frac{\left(c\tau+d\right)^{3} }{((c\tau +d)m+n)^3}\right)
\eea  
showing  the obstruction to modularity.

\section{The D(-1)D7 system on $\mathbb{R}^4\times \mathbb{R}^4/\mathbb{Z}_2$}

\begin{table}[t]
	\begin{center}
		\begin{tabular}{|c|c|c|c|}
			\hline
			$(\Phi,{\cal Q} \Phi )$ & $(-)^F$ &$U(k)\times U(N)\times U(M)$ & ${\cal Q}^2$ \\
			\hline
			$(B_\ell,M_\ell )$ & $ + $ &  $  ({\bf k} {\bf \bar k} , {\bf \one}, {\bf \one} )$ & $T_\ell \, {x_i\over x_j} $ \\
			$(\lambda^{IJ},D^{IJ} )$ & $-$ & $  ({\bf k} {\bf \bar k} , {\bf \one}, {\bf \one} )$ & $T_I T_J\, {x_i\over x_j} $ \\
			$(\lambda_+,D_7 )$ & $-$ & $  ({\bf k} {\bf \bar k} , {\bf \one}, {\bf \one} )$ & $  {x_i\over x_j} $ \\
			$(\bar\chi,\lambda_- )$ & $+$ & $  ({\bf k} {\bf \bar k} , {\bf \one}, {\bf \one} )$ & $ {x_i\over x_j} $ \\
			$(w,\mu)$ & $+$ & $  ({\bf \bar k}  , {\bf  N}, {\bf \one} )$ & ${T_a\over x_i} $ \\
			$(\mu', h')$ & $-$ & $  ({\bf \bar k} , {\bf \one} , {\bf M} )$ & ${T_m\over x_i} $ \\
			\hline
		\end{tabular}
	\end{center}
	\caption{Instanton moduli space. The domain of the indices for
		$\mathbb{R}^4\times \mathbb{R}^4/\mathbb{Z}_2$ are: 
		$\ell,I,J=1, 2$ with $I<J$, $i, j=1,\ldots,k$, $u=1,\ldots,N$ and $f=1,\ldots,M$.   }
	\label{tablemoduli4}
\end{table}% 

As we stated in the introduction the eight dimensional case is a clean example in which the non perturbative effects due to what we called exotic instantons are displayed. Nonetheless, possible phenomenological applications
should take place in our four dimensional world. It is possible to compute such effects wrapping
N fractional D7 and M fractional D7' branes on $\mathbb{R}^4/\mathbb{Z}_2$ and focusing on the exotic prepotentials generated by the D(-1)-instantons. The prepotential obtained in such a way defines a four dimensional effective action
\be
S_{\rm eff}=\int d^4xd^4\theta\, {\cal F}(\Phi,W) %=\tau(a,\epsilon)\int \,d^4x \,tr F^2+\ldots
\label{effaction}\eeq
where $\Phi, W$ are the superfields of which the scalar field and the $\Omega$ background are one of the components .
%and $\tau(A)$ is the second derivative of the prepotential.
The form (\ref{effaction}) is dictated by supersymmetry, the important difference with the standard case is that prepotentials generated by
exotic instantons are polynomials in the vev. These effects are purely stringy, generated by wrapped euclidean branes along the internal space  and not present in "standard" field theory. 
%A model containing both instanton contributions could be easily cooked up in a quiver type theory in which the total gauge group is the product of the gauge groups containing separately "standard" and exotic instantons.
The moduli content of the exotic theory is given in Table
\ref{tablemoduli4}. With respect to the eight dimensional case, the fields labeled by the indices $\ell=3,4$ and $I,J=3$ are projected out by the orbifold quotient. The character reduces to
\bea 
\label{charorb} 
\chi ={\rm tr}_{\cal H}  e^{-\beta {\cal Q}^2 } ={-}\sum_{i,j=1}^k\frac{x_i}{x_j} (1-T_1)(1-T_2)+\sum_{i=1}^k \left( \sum_{u=1}^N { T_{a_u}  \over x_i } {-}   \sum_{f=1}^M   { T_{m_f}  \over x_i}  \right) 
\eea
With respect to the nine dimensional character (\ref{char}) there is no need for a condition of the type $i<j$ given that a condition of the type of (\ref{epsilonconstraint9d}) is absent and the "volume" $(1-T_1)(1-T_2)$
correctly contains the four terms corresponding to the degrees of freedom of the fields $B_1, B_2$ and the constraints due to the $U(k)$ symmetry and the complex ADHM constraint.
Following the same procedure of the previous section we find 
\bea
Z_k &=&    {{\cal E}^k\over k!} \int  \prod_{i=1}^k {dx_i\over x_i}  \prod_{i \neq j}^k\frac{ \left(1-{x_i \over x_j} \right)  \left( 1-{x_i \over x_j} T_1 T_2 \right)   }{  \left( 1-{x_i \over x_j} T_1\right)\left( 1-{x_i \over x_j} T_2\right) }    \prod_{i=1}^k \frac{\prod_{f=1}^M \left(1-{T_{M_f} \over x_i} \right)} {\prod_{u=1}^N  \left(1-{T_{a_u} \over x_i} \right)  }
\label{localization1}
\eea
with
\be
{\cal E}= \frac{ \left( 1-  T_1 T_2 \right)  }{  \left( 1-  T_1\right)\left( 1-  T_2\right) }
\eeq
The partition function can now be written in the plethystic form
\be
Z_{5d}={\rm exp}\left[\sum_{n=1}^\infty { f'(T^n_\ell, T^n_\mu,q^n) \over n } \right]
\eeq
with 
\bea
&&f'(T_\ell, T_\mu,q)=\frac{(1-T_1T_2)(1-T_{\mu})q}{(1-T_1)(1-T_2)}\qquad {\text if}\quad M=N\\
&&f'(T_\ell, T_\mu,q)=\frac{(1-T_1T_2)q}{(1-T_1)(1-T_2)}\hspace{1.4cm} {\text if}\quad M=N-1\nn\\
&&f'(T_\ell, T_\mu,q)=0\hspace{4cm} {\text if}\quad M\le N-2\nn
\eea 
 Evaluating the first few instanton orders and sending $\beta, \epsilon_\ell \to 0$, one finds that the four dimensional prepotential 
 resums to the simple form
 \bea
&&{\cal F}_4= -\mu (\epsilon_1+\epsilon_2) \log(1- q ) \qquad {\text if}\quad M=N\\
&&{\cal F}_4= (\epsilon_1+\epsilon_2) q \hspace{2.8cm} {\text if}\quad M=N-1\nn\\
&&{\cal F}_4=0 \hspace{4.2cm} {\text if}\quad M\le N-2\nn
 \eea
 with $\mu$ given again by (\ref{mu}). 
Also in this four dimensional case we find that only the $U(1)$ gauge degrees of freedom correct the classical prepotential.

\subsection{Chiral Correlators}

Chiral correlators $\langle {\rm tr}\phi^J \rangle$ in the four-dimensional  gauge theory are computed now by  
\be
\label{trfijorb}
\langle {\rm tr}\, e^{ z\, \phi} \, \rangle = {\rm tr}\, e^{ z\, a}- {1\over Z_{4d} }  \sum_{Y} q^{|Y|}\, Z_{4Y}    \sum_{i=1}^k \left(x_{i}^Y \right)^z (1-T_1^z) (1-T_2^z) 
\eeq
with $Z_Y$ the contribution to the partition function of the solid partition set $Y=\{ Y_u \}$  and $x_{i}^Y$ denotes the location of the corresponding pole given by (\ref{Vspace}).  We notice that in the four-dimensional limit $\beta \to 0$, the right hand side of (\ref{trfijorb}) starts as $z^2$, so all correlators with $J<2$ receive no instanton corrections. The first non-trivial correlator is then given by the coefficient of the term $z^2$ 
in (\ref{trfijorb})
\be
\label{trfij2}
\langle {\rm tr}\,   \phi^2 \, \rangle = {\rm tr}\, a^2- {2\epsilon_1\epsilon_2  \over Z_{4d} }  \sum_{Y} q^k\, k Z_{4k}   
=   {\rm tr}\, a^2 -2q{d\over dq} {\cal F}_4(q)=  {\rm tr}\, a^2-2\mu(\epsilon_1+\epsilon_2)\,{q \over 1-q }
\eeq
  Remarkably, as in the eight dimensional case, the chiral correlators of the orbifold theory,
at the leading order in the $\Omega$ background, can be represented in a closed form. The result can be written as
\bea 
\label{resolv_exact_mass_orb} 
\left\langle tr\,\frac{1}{x-\phi} \right\rangle =\frac{P'(x)}{P(x)}
+(\epsilon_1+\epsilon_2)\frac{\partial^2}{\partial x^2}\left[\log 
\left(\frac{1-q}{ \left(1-q R(x)\right)}\right)+O(\epsilon)\right]\,,
\eea 
where $P(x)$, $R(x)$ are given again by (\ref{P_Q_def}). One finds again exotic instanton corrections to the correlators in the SU(N) theory.
For example for the $SU(2)$ theory, with two flavours and $a_1=-a_2=a$, $m_1=-m_2=m$, the first non-trivial correlators are 
\bea
\langle {\rm tr}\,   \phi^3 \, \rangle &=& 6 ( \epsilon_1+\epsilon_2) (a^2-m^2) ( q+ q^2+q^3+\ldots )\nn\\
 \langle {\rm tr}\,   \phi^5 \, \rangle &=& 10 ( \epsilon_1+\epsilon_2) (a^2-m^2) ( 2 a^2 q+  (3a^2-m^2) q^2+ 2 (2a^2-m^2) q^3+\ldots ) 
\eea
and so on.

\vskip 0.6cm
\noindent {\large {\bf Acknowledgments}}
\vskip 0.02cm
R.Poghossian wants to thank the I.N.F.N. unit Tor Vergata for hospitality and financial support during the completion of this work.  He also acknowledges the support in the framework of Armenian SCS grant 20RF-142.  
\appendix
\section{Transformation properties of the function $D_{s}(\tau)$}

In this appendix we study the transformation properties of the series
\be
D_s(\tau)=\sum_{n=1}^\infty \sigma_{s-1}(n)q^n=	\sum_{n=1}^\infty\frac{n^{s-1}q^n}{1-q^n}
\label{lambertseries}\eeq
where $\sigma_{s}(n)=\sum_{d|n}d^s$ is the sum over the positive divisor function, under the modular group 
\be
\tau\to\frac{a\tau+b}{c\tau+d}, \quad  \begin{pmatrix}
	a & b\\
	c&d 
\end{pmatrix}\in SL(2,\Z)
\label{sl2z}\eeq
An integral formula for these transformations is given in \cite{Kim:2019}
\bea
&&D_{3}\left(\frac{a\tau+b}
{c\tau+d}\right)=\left(c\tau+d\right)^{s}\left[ D_{s}(\tau)+\frac{1}{2}\left(1-\left(c\tau+d\right)^{-s}\zeta(3)\right)\right.\nonumber\\
&+&\left.\frac{c^{s-1} sec\frac{\pi s}{2}}{8\left(c\tau+d\right)^{s/2}}\int_C(-z)^{s-1}\left( c+\sum_{j=0}^{c-1}cot\pi\left(i\sqrt{c\tau+d }z-\frac{j d}{c}\right)cot\pi\left(\frac{i z}{\sqrt{c\tau+d }}-\frac{j}{c}\right)\right)dz \right]\nonumber
\label{formulakim}\eea
where C is the Hankel contour encircling the positive real axis in the clockwise direction but not other poles of the integrand. 
From its definition (\ref{lambertseries}), it is clear that $D_s(\tau)$ is invariant under T: $\tau\rightarrow \tau+1$ transformations. The
S: $\tau\rightarrow -1/\tau$ transformation yields
\bea
D_s\left(-\frac{1}{\tau}\right)&=&\tau^s \left(D_s(\tau)+\ft{1}{2}(1-\tau^{-s}\zeta(s-1))
\right.\nonumber\\
&+&\left.\frac{1}{8\tau^{s/2}\cos (\pi s/2)}
\int_C (-z)^{s-1}\left(1+\cot\left(\pi i \sqrt{\tau} z\right)\cot\frac{\pi iz}
{ \sqrt{\tau}}\right)dz\right)
\label{kimformulas3}\eea
For $s=3$ the second line in (\ref{kimformulas3}) is  an indeterminate form. Using L'Hopital rule we get
\bea
\label{int2}
\frac{1}{4\pi \tau^{3/2}}
\int_C z^2 \log (-z)\left(1+\cot\left(\pi i \sqrt{\tau} z\right)\cot\frac{\pi iz}
{ \sqrt{\tau}}\right)dz
\eea
 The values of logarithm along the upper and lower branches of the contour $C$ differ by $2\pi i$, so the contour 
integral (\ref{int2}) can be rewritten as an ordinary integral
\bea
\label{int3}
\frac{i}{2\tau^{3/2}}
\int_0^\infty x^2\left(1+\cot\left(\pi i \sqrt{\tau} x\right)\cot\frac{\pi ix}
{ \sqrt{\tau}}\right)dx
\eea 
or equivalently
\bea
\label{int4}
-\frac{i}{\tau^{3/2}}
\int_0^\infty x^2\,\,\frac{e^{-\frac{2 \pi  x}
{\sqrt{\tau }}}+e^{-2 \pi  \sqrt{\tau } x}}
{\left(1-e^{-\frac{2 \pi  x}{\sqrt{\tau }}}\right) 
\left(1-e^{-2 \pi  \sqrt{\tau } x}\right)}\,\,dx
\eea
Expanding the denominator as a double geometric series and performing the elementary integration we obtain
\bea
\frac{i}{2\pi^3}\left(\sum_{m,n=1}^\infty\frac{1}{(m \tau +n)^3}
+\frac{1+\tau^{-3}}{2}\zeta(3)\right)
\eea
 leading to
\bea 
\label{S_transform}
\tau^{-3}D_3(-1/\tau)=D_3(\tau)+
\frac{i}{2\pi^3}\left(\sum_{m,n=1}^\infty\frac{1}{(m \tau +n)^3}
+\frac{1+\tau^{-3}}{2}\zeta(3)\right)
\eea 
which already appeared in \cite{shimomura}. An equivalent formula for the S-transformation can be obtained by using the T-invariance and considering the modular transformation 
\be
TST : \quad  \tau \to -{1\over \tau+1}+1 =-{1\over  -1-\tau^{-1} } 
\eeq
 One finds
\bea 
\label{TSTS_transform}
&&-\tau^{-3}D_3(-1/\tau)=D_3(\tau)\nonumber\\
&&+\frac{i}{2\pi^3}\left(\sum_{m,n=1}^\infty\frac{1}{(m (\tau+1)+n)^3}
+\frac{1}{(-m (\tau+1) +n\tau)^3}
+\frac{1+\tau^{-3}}{2}\zeta(3)\right)
\eea  
Summing (\ref{S_transform}) with (\ref{TSTS_transform}) we see that the l.h.s.'s 
cancel and the double sums and $\zeta(3)$  
nicely combine into
\bea
\label{D2_double_sum}
D_3(\tau)=-\frac{i}{4\pi^3}\sum_{m=1}^\infty
\sum_{n=-\infty}^\infty\frac{1}{(m \tau +n)^3}
\eea
Similar manipulations lead to the general double sum representation  \cite{shimomura}
\bea
\label{Ds_double_sum}
D_{s}(\tau)=\frac{e^{\pi i s/2}\Gamma(s)}{(2\pi)^s}\sum_{m=1}^\infty
\sum_{n=-\infty}^\infty\frac{1}{(m \tau +n)^s}
\eea
from which it follows 
\bea 
D_{s}\left(\frac{a\tau+b}{c\tau+d}\right)=\frac{e^{\pi i s/2}
	\Gamma(s)}{(2\pi)^s}\sum_{m=1}^\infty
\sum_{n=-\infty}^\infty\frac{(c\tau+d)^s}{(m (a\tau+b)+n(c\tau+d))^s}\nonumber\\
=\frac{e^{\pi i s/2}
	\Gamma(s)(c\tau+d)^s}{(2\pi)^s}\sum_{m=1}^\infty
\sum_{n=-\infty}^\infty\frac{1}{((m a+n c)\tau+(m b+n d))^s}\nonumber\\
=\frac{e^{\pi i s/2}
	\Gamma(s)(c\tau+d)^s}{(2\pi)^s}\sum_{({\tilde n},{\tilde m})\in \Omega}
\frac{1}{({\tilde m}\tau+{\tilde n})^s}
\label{dsformula}\eea

\begin{figure}
	\centering 
	\begin{tikzpicture}[scale=0.50]
	\coordinate (Origin)   at (0,0);
	\coordinate (XAxisMin) at (-7.5,0);
	\coordinate (XAxisMax) at (7.5,0);
	\draw [thin, gray,
	%-latex
	] (XAxisMin) -- (XAxisMax);% Draw x axis	
	%\clip (-6,-6) rectangle (12cm,12cm); % Clips the picture...
	\foreach \x in {-7,...,7}{% Two indices running over each
		\foreach \y in {-5,...,5}{% node on the grid we have drawn 
			\node[draw,circle,inner sep=1.5pt] at (\x,\y) {};
			% Places a dot at those points
		}
	}
	\node at (0,-0.47){\footnotesize{0}};
	\node at (0,1-0.45){$\tau$};
	\node at (1,-0.47){\footnotesize{1}};
	\draw[thin,red] (-7.5,-15/3)--(7.5,15/3);
	\foreach \x in {-7,...,-1}{
		\node[draw,red,circle,inner sep=1.5pt,fill] at (\x,0) {};}
	\foreach \x in {-7,...,-2}{
		\node[draw,red,circle,inner sep=1.5pt,fill] at (\x,-1) {};}
	\foreach \x in {-7,...,-4}{
		\node[draw,red,circle,inner sep=1.5pt,fill] at (\x,-2) {};}
	\foreach \x in {-7,...,-5}{
		\node[draw,red,circle,inner sep=1.5pt,fill] at (\x,-3) {};}
	\foreach \x in {-7,...,-7}{
		\node[draw,red,circle,inner sep=1.5pt,fill] at (\x,-4) {};}
	\foreach \x in {2,...,7}{
		\node[draw,circle,inner sep=1.5pt,fill] at (\x,1) {};}
	\foreach \x in {3,...,7}{
		\node[draw,circle,inner sep=1.5pt,fill] at (\x,2) {};}
	\foreach \x in {5,...,7}{
		\node[draw,circle,inner sep=1.5pt,fill] at (\x,3) {};}
	\foreach \x in {6,...,7}{
		\node[draw,circle,inner sep=1.5pt,fill] at (\x,4) {};}
	\end{tikzpicture}
	\caption{Summation regions. The case $a=3$, $b=4$, $c=2$, $d=3$}
	\label{sumregions}
\end{figure}
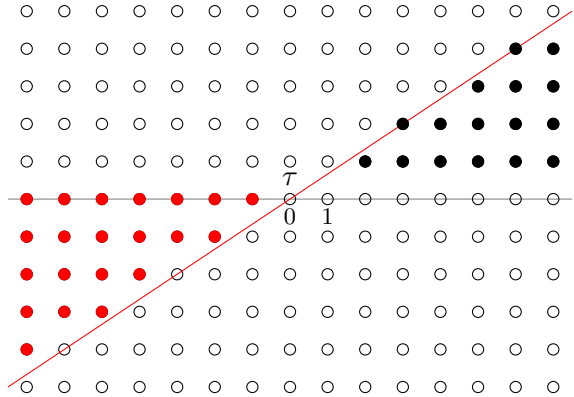

%%%%%%%%%%%%%%%%%%%%%%%%%%%%%%%%%%%%%%%%%%%%%%%%%%%%%%%%%%%%%%%%%%%%%%%%%%%
  The sum in the last line runs over the points above the line $L=\{(n,m)|cn-dm=0\}$ (the red line in Fig.\ref{sumregions}). Comparing against the sum in (\ref{Ds_double_sum}) running over points in the upper half plane one finds that in the difference
\[
(c\tau+d)^{-s}D_{s}\left(\frac{a\tau+b}{c\tau+d}\right)-D_{s}\left(\tau\right)
\]
all the white dots in Fig.\ref{sumregions} cancel and one is left with a sum over the red dots 
minus a sum over the black dots. Now it becomes obvious 
the crucial difference between the cases with even and odd values of $s$. In 
the former case the contributions of the red and black dots (besides those which lie 
on the black and red lines respectively) completely cancel out and we get 
\[
\left(c\tau+d\right)^{-s}\left(\frac{e^{\pi i s/2}
	\Gamma(s)\zeta(s)}{(2\pi)^s}+D_{s}\left(\frac{a\tau+b}
{c\tau+d}\right)\right)=\frac{e^{\pi i s/2}
	\Gamma(s)\zeta(s)}{(2\pi)^s}+D_{s}\left(\tau\right)
\]
where the $\zeta$ functions are just the contributions of the dots 
lying on the black and red lines. As a result the combination
\[
\frac{e^{\pi i s/2}
	\Gamma(s)\zeta(s)}{(2\pi)^s}+D_{s}\left(\tau\right)
\] 
is a weight $s$ modular form. Instead for odd $s$
we get twice the contributions of the inner  red boxes (those lying 
strictly below the black line) together with the difference of 
the contributions of the red dots lying on the black line and the black dots 
lying on the red line. The final result now reads :  
\bea
&&\left(c\tau+d\right)^{-s}D_{s}\left(\frac{a\tau+b}
{c\tau+d}\right)=D_{s}\left(\tau\right)\nonumber\\
&&-\frac{e^{\pi i s/2}
	\Gamma(s)}{(2\pi)^s}\left(\left(1+(c\tau+d)^{-s}\right)
\zeta(s)+2\sum_{m=1}^\infty\sum_{n=1}^\infty 
\frac{1}{((c\tau +d)m+n)^s}\right)
\eea 
In particular for $s=3$ we find the transformation (\ref{d3transf}).  
 
\providecommand{\href}[2]{#2}\begingroup\raggedright\endgroup

\end{document}